\documentclass[lettersize,journal]{IEEEtran}
\usepackage{enumitem}
\usepackage{url}
\usepackage{graphicx}
\usepackage{multirow}
\usepackage{comment}
\usepackage{xcolor}
\usepackage{textcomp}
\usepackage{dblfloatfix}    
\usepackage{lipsum}
\usepackage{scrextend}
\usepackage{comment}
\usepackage{amsmath,amssymb}
\usepackage{hhline}
\usepackage{tabularx,ragged2e}
\usepackage{caption}
\usepackage{subcaption}
\usepackage{cite}
\usepackage{lipsum}
\usepackage{ragged2e}
\usepackage[hidelinks]{hyperref}
\usepackage{tikz}
\usepackage{cite}
\usepackage{esvect}
\usepackage{algorithm}
\usepackage[noend]{algpseudocode}
\usepackage{mathtools}
\usepackage{verbatim}
\newcommand*\circled[1]{\tikz[baseline=(char.base)]{
            \node[shape=circle,draw,inner sep=2pt] (char) {#1};}}
\newcolumntype{C}{>{\Centering\arraybackslash}X}

\newcommand{\name}[1]{{AIRIC}}
\newcommand{\vrain}[1]{{TMC'20}}
\newcommand{\rev}[1]{{\color{black}{#1}}}

\usepackage{calc}
\newlength\myheight
\newlength\mydepth
\settototalheight\myheight{Xygp}
\algdef{SE}[DOWHILE]{Do}{doWhile}{\algorithmicdo}[1]{\algorithmicwhile\ #1}

\DeclareMathOperator{\EX}{\mathbb{E}}

\DeclareMathOperator*{\argmin}{arg\,min}

\ExplSyntaxOn
\NewDocumentCommand{\prow}{m}
 {
  \seq_set_split:Nnn \l_tmpa_seq { , } { #1 }
  \seq_use:Nn \l_tmpa_seq { & }
  \\
 }
\ExplSyntaxOff

\begin{document}

\title{AIRIC: Orchestration of Virtualized Radio Access Networks with Noisy Neighbours\\
{\large As accepted in the IEEE Journal on Selected Areas in Communications 2023}
}

\author{J. Xavier Salvat Lozano,~\IEEEmembership{Member,~IEEE,} Andres Garcia-Saavedra, Xi Li, Xavier Costa Perez,~\IEEEmembership{Senior Member,~IEEE}
\thanks{J. Xavier Salvat Lozano, A. Garcia-Saavedra and Xi Li are with NEC Laboratories Europe GmbH, Heidelberg, Germany (e-mails: \{name.surname\}@neclab.eu).}
\thanks{X. Costa-Pérez is with NEC Laboratories Europe GmbH, Heidelberg, Germany, and i2CAT Foundation and ICREA, Barcelona, Spain (e-mail: xavier.costa@ieee.org).}
\thanks{The work was supported by the European Commission through Grants No.
SNS-JU-101097083 (BeGREEN) and 101017109 (DAEMON). Additionally,
it has been supported by MINECO/NG EU (No. TSI-063000-2021-7) and the
CERCA Programme.}}

\markboth{Journal on Selected Areas in Communications}{}

\maketitle

\IEEEpubid{\begin{minipage}{\textwidth}\ \\[12pt]
\copyright2023 IEEE.  Personal use of this material is permitted.  Permission from IEEE must be obtained for all other uses, in any current or future media, including reprinting/republishing this material for advertising or promotional purposes, creating new collective works, for resale or redistribution to servers or lists, or reuse of any copyrighted component of this work in other works.
\end{minipage}}

\IEEEpubidadjcol

\begin{abstract}
Radio Access Networks virtualization (vRAN) is on its way becoming a reality driven by the new requirements in mobile networks, such as scalability and cost reduction. Unfortunately, there is no free lunch but a high price to be paid in terms of computing overhead introduced by \rev{\emph{noisy neighbors} problem} when multiple virtualized base station instances share computing platforms. In this paper, first, we thoroughly dissect the multiple sources of computing overhead in a vRAN, quantifying their different contributions to the overall performance degradation. Second, we design an \emph{AI-driven Radio Intelligent Controller (\name{})} to orchestrate vRAN computing resources. \name{} relies upon a hybrid neural network architecture combining a relation network (RN) and a deep \rev{Q-Network} (DQN) such that: ($i$) the demand of concurrent virtual base stations is satisfied considering the overhead \rev{posed by the \emph{noisy neighbors} problem} while the operating costs of the vRAN infrastructure is minimized; and ($ii$) dynamically changing contexts in terms of network demand, signal-to-noise ratio (SNR) and the number of base station instances are efficiently supported. Our results show that \name{} performs very closely to an offline optimal oracle, attaining up to $30\%$ resource savings, and substantially outperforms existing benchmarks in  service guarantees. 
\end{abstract}


\IEEEpubidadjcol

\section{Introduction} \label{sec:intro}

Radio Access Network (RAN) virtualization is well-recognized as a key technology to increase cost-efficiency at the very edge of next-generation mobile systems~\cite{samsung_whitepaper}. The urge to increase the density of radio access points---yet preserve or even reduce costs---has attracted the attention of industry in this direction; see, e.g., initiatives such as the O-RAN alliance~\cite{garcia2021ran} or Rakuten's greenfield deployment in Japan~\cite{rakuten_whitepaper}. 
Virtualized RANs (vRANs) are expected to import the advantages of NFV such as resource multiplexing by sharing infrastructure~\cite{nuberu}. The idea of RAN pooling is not new: 71\% of US operators indicated the intent to deploy RAN centralization by 2025 in a recent survey~\cite{cran-survey}, e.g., NTT Docomo, Ericsson or AT\&T are famously interested this type of technologies~\cite{ntt-centralization, ericsson-centralization, att-centralization}; and centralization is at the forefront of O-RAN~\cite[\S 5.1.3]{oran-cloud-scenarios}. However, the real-time impact of resource contention in shared RAN pooling platforms has not been studied sufficiently.

The success of Network Function Virtualization (NFV) has spurred the market to build virtual network functions (VNFs) such as firewalls, switches, VPNs, etc., that provide carrier-grade performance.
However, research has shown that resource contention caused by VNFs sharing common computing infrastructure may lead to up to 40\% of performance degradation compared to dedicated platforms~\cite{tootoonchian2018resq, manousis2020contention}. The term \emph{noisy neighbor} problem has been coined to refer to this issue, and has motivated substantial research over the years~\cite{sun2017nfp, kumar2019picnic, manousis2020contention, gong2020microscope, tootoonchian2018resq}. See our review on the related work in \S\ref{sec:related}.

The virtualization of base stations (vBSs) is not alien to this issue. We confirm this with our own findings from experiments in a proof-of-concept vRAN system comprised of instances of a full-fledged 3GPP Rel.10 compliant vBS implemented with {\ttfamily srsRAN}~\cite{srslte}. Using Docker container techniques, we deployed a set of 10MHz vBS instances in a pool of CPU cores from an Intel core i7-7700K CPU @ 4.20GHz in a shared off-the-shelf server. The details of our experimental setup will be presented later. We then initiated bidirectional data flows, both uplink (UL) and downlink (DL), with maximum load and good wireless channel conditions between each vBS instance and a corresponding legacy user equipment (UE). 

\begin{figure}[t!]
 \centering
\minipage{0.47\columnwidth}
 \includegraphics[width=\columnwidth]{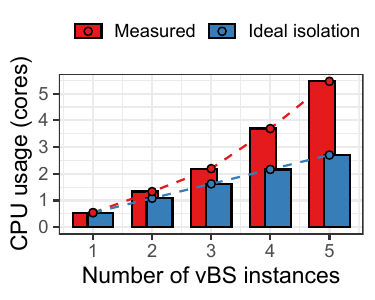}
\vspace{-5mm}
\caption{\small vRAN per-core CPU usage with \# of vBS.}
 \label{fig:intro:vran}
\endminipage{}
\hfill
\minipage{0.47\columnwidth}
\includegraphics[width=\columnwidth]{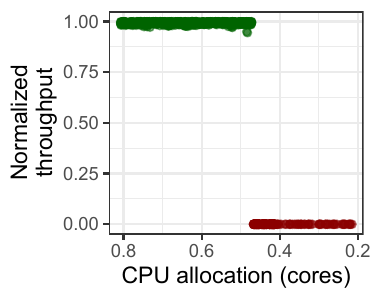}
\vspace{-5mm}
\caption{\small Throughput \emph{vs.} CPU allocation.}
 \label{fig:intro:thr_vs_cpualloc}
\endminipage{}
\vspace{-5mm}
\end{figure}

\IEEEpubidadjcol

Fig.~\ref{fig:intro:vran} depicts the relative CPU usage of the system as a function of the number of vBS instances deployed. The bars in blue show the expected usage assuming perfect resource isolation in place. We compute these by linearly scaling up the CPU usage of a single vBS instance. The red bars show the actual CPU consumption, which unveil an exponentially-growing overhead induced by the aforementioned resource contention in imperfectly isolated computing platforms.

In the context of vRAN, exploring the gains and impact of radio network function virtualization may prove challenging to consider RAN specific characteristics. First, the vBS workload has strict time deadlines, which makes them much more sensitive to \rev{the \emph{noisy neighbors} problem} than classical VNFs such as switches or firewalls. We confirm this in Fig.~\ref{fig:intro:thr_vs_cpualloc}, which shows the normalized throughput performance of one vBS for different CPU allocations (x-axis). Note that its throughput rapidly collapses upon deficit of computing resources. This occurs because physical layer (PHY) deadlines are missed, which causes that users lose synchronization with the vBS, resulting in connectivity loss~\cite{nuberu}. This differs significantly from the cases of regular VNFs, which suffer from a smoother performance degradation upon computing resource shortages. Hence, it is an essential problem to compute required shared computing resources for vRAN deployments accounting for such impact of \rev{the \emph{noisy neighbour} problem}, which is the aim of this work. 

vRANs inspired remarkable work over the last few years. In the industry, Intel FlexRAN and NVIDIA Aerial are vRAN solutions that use dedicated hardware accelerators, which are overly expensive and energy-consuming~\cite{nuberu}. In the academia, Agora~\cite{agora} proved that RAN PHY tasks can be executed in many-core general-purpose CPU platforms with carrier-grade performance, but it requires  CPU cores to be \emph{dedicated} to specific tasks (i.e., no sharing). More recently, Concordia~\cite{concordia} proposed an approach to share computing resources with latency-elastic applications. However, \emph{how to share computing resources across several vBSs remains an open question}.

Although much work has studied \rev{the noisy neighbours problem} on NFV workloads~\cite{tootoonchian2018resq}, little research has been done on the vRAN case. Nuberu~\cite{nuberu} provides a RAN PHY processing pipeline that increases its reliability upon computing capacity fluctuations but it does not deal with the CPU allocation problem. vrAIn~\cite{vrain, vrain2} does address this problem but it does not consider the impact of 
 \rev{the noisy neighbors problem} (perfect resource isolation is assumed) and it does not support a variable number of vBSs in the system (see \S\ref{sec:evaluation}). To the best of our knowledge, we are the first to address the vRAN noisy neighbor problem on shared computing platforms (see related work in \S\ref{sec:related}). More specifically, we provide the following contributions:
\begin{itemize}[noitemsep,topsep=0pt,parsep=0pt,partopsep=0pt,leftmargin=8pt]
\item In \S\ref{sec:dissection}, we provide an in-depth analysis of the overhead incurred by multiple vBSs sharing a common CPU pool.
 \item In \S\ref{sec:design}, we design a data-driven model called \name{} to optimize the allocation of computing resources in a vRAN. Compared to state-of-the art solutions, our approach learns to compensate for the overhead caused by resource contention and supports a varying number of vBS instances without requiring independent models.
  \item In \S\ref{sec:evaluation}, we empirically compare \name{} with related solutions~\cite{vrain, vrain2} and with an optimal offline oracle. We show that \name{} achieves close-to-optimal performance and over 99.9\% throughput service. In contrast, previous solutions provide barely $7\%$ savings in computing resources at a price of up to $50\%$ throughput loss.
\end{itemize}

\section{Background}\label{sec:background}

\subsection{Radio Access Network Virtualization}

\begin{figure}[t!]
 \centering
 \includegraphics[width=\columnwidth]{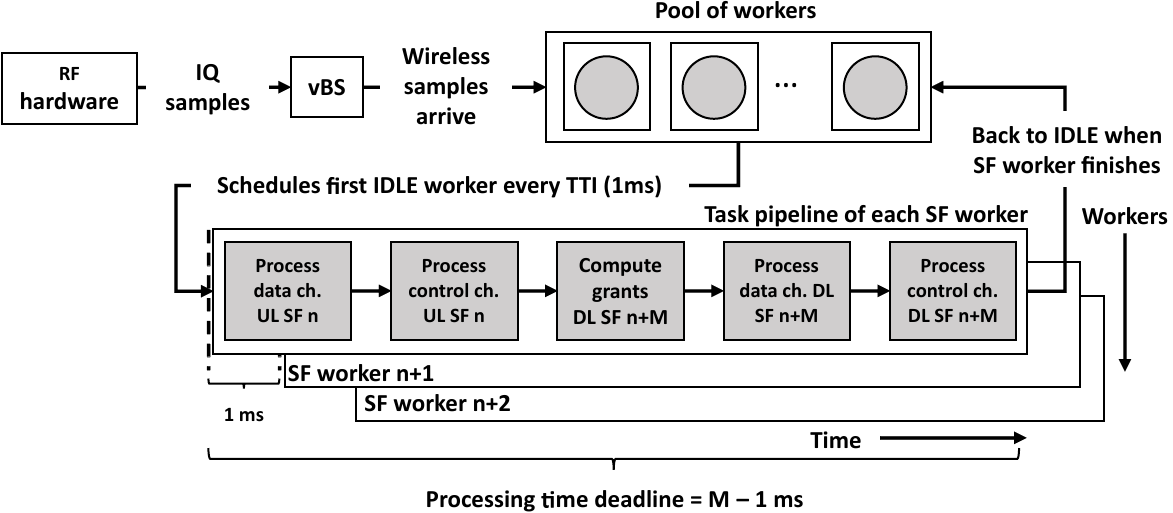}
 \vspace{-6mm}
 \caption{\small Every TTI a vBS needs to spawn a new thread for its pool to process the different tasks for UL and DL}
 \label{fig:bk:vbs_thr}
 \vspace{-4mm}
\end{figure}

It is well-known that the physical layer (PHY) of a vBS stack carries most of the computing heavy-lifting~\cite{7423494}. We next provide some background about this.
Fig.~\ref{fig:bk:vbs_thr} illustrates the operation of a Frequency Division Duplex (FDD) vBS PHY processor~\cite{nuberu}. 

Every 1 ms, a vBS receives the radio samples associated with an uplink subframe $n$. A dispatcher selects an idle worker, which initiates a pipeline of radio processing tasks in an independent computing thread. These tasks include ($i$) processing the data and control channels carried by the UL subframe $n$, ($ii$) scheduling UL/DL radio grants to be transported by DL subframe $n + M$, ($iii$) processing data and control channels for DL subframe $n + M$, and ($iv$) send the modulated symbols corresponding to DL subframe $n + M$ to the radio frontend. 
In 4G LTE, $M=4$ in respect to 3GPP constraints to provide hybrid ARQ feedback to the users, but this parameter is configurable in 5G New Radio~\cite{3gpp38211}. 

Processing channels in a subframe consists of additional pipelines of operations, including (de)modulation of OFDM symbols or forward error coding (FEC) operations, which are compute-intensive. However, a downlink subframe has to be generated every 1~ms, and an uplink subframe has to be processed every 1~ms. To give the worker some slack to execute its job, pipeline parallelization is used. That is, a pool of $M-1$ workers shall be available to execute jobs. Once a worker finishes a job, it becomes idle awaiting new jobs.  

Among all virtualization technologies available today (virtual machines, unikernels, containers), we believe that Docker containers are the best fit to support the requirements of vRAN workloads~\cite{garcia2021ran}.
To begin with, Docker containers support online granular resource allocation and orchestration of multiple tenants across multiple hosts. Furthermore, as opposed to virtual machines, Docker supports fast and live migration of containers, as well easy and quick creation, upgrade, and deployment of images. 

\subsection{General-purpose Computing}

\begin{figure}[t!]
 \centering
 \includegraphics[width=0.7\columnwidth]{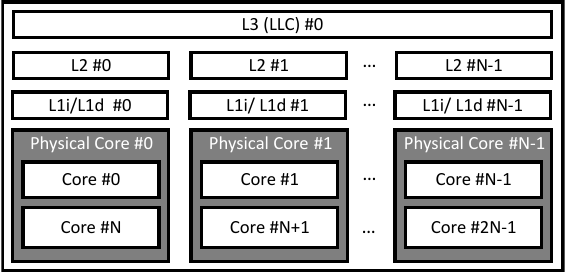}
 \vspace{-1mm}
 \caption{\small General-purpose CPU platform architecture.}
 \label{fig:bkgrd:cpu_topo_testbed}
\vspace{-5mm}
\end{figure}

Fig. ~\ref{fig:bkgrd:cpu_topo_testbed} presents the CPU architecture of a general-purpose computing platform (GPP). Modern \rev{superscalar} processors leverage on simultaneous \rev{multithreading} (SMT) (also known as \rev{Hyper-threading} in Intel CPUs), which allows a physical core to run more than one thread at time. Thus, physical cores are seen from the operating system as two separated cores. These cores are virtual and share the same physical processor.

The cache memory is the closest and fastest memory of the CPU. It bridges the gap between RAM memory speed and CPU speed. Cache memory is usually organized in different levels regarding speed and size~\cite{jacob2010memory}. Level 1 (L1) cache memory is the closest and fastest memory of the system but also its capacity is the most limited. Each physical core has its dedicated L1 cache. L2 cache is bigger than L1 but slower, and it is also dedicated to each physical core. As opposed to L1, L2 is generally used for data rather than instructions. Finally, L3 cache or Last Level Cache (LLC) is the slowest cache of a CPU, and it is shared across all cores.

In a GPP, a core executing a thread loads the most used memory blocks into a cache  for faster access. Then, every time a thread references a memory block that is not in a cache, the core triggers an interrupt called a ``cache miss'', and looks for the data in a  higher-layer memory cache (or RAM, ultimately).

\subsection{O-RAN architecture} \label{sec:oran_archl}

\begin{figure}[t!]
 \centering
 \includegraphics[width=0.7\columnwidth]{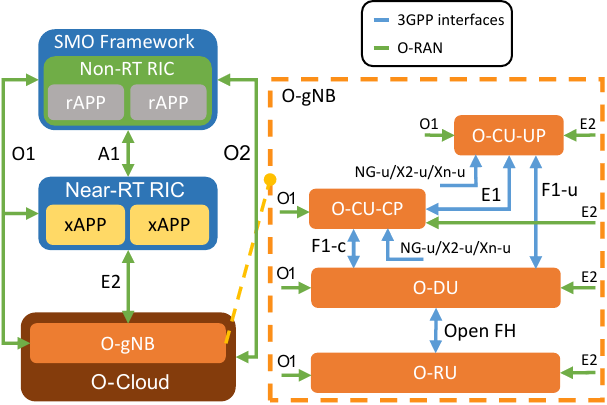}
 \vspace{-1mm}
 \caption{\rev{\small O-RAN architecture.}}
 \label{fig:bkgrd:oran_arch}
\vspace{-4mm}
\end{figure}

The O-RAN alliance \cite{9579445} is a joint collaboration between leading industry and carrier partners in the mobile communications sector to redesign future Radio Access Network (RAN) technologies. Its main goal is to define a technical standard for RAN architecture that fosters innovation, interface openness, and reducing operational and deployment costs thanks to virtualization and general-purpose hardware.

Fig.~\ref{fig:bkgrd:oran_arch} depicts the general outlook of the O-RAN architecture. O-RAN splits the BS functions into three Network Functions (NFs): ($i$) a Radio Unit (O-RU), ($ii$) a Distributed Unit (O-DU), and  ($iii$) a Central Unit (O-CU)~\cite{oranwg1}. The O-RU hosts low-level PHY functions, including FFT and other RF functions such as amplification or sampling. The O-DU hosts the RLC, MAC and high PHY layers, which include FEC encoding and decoding. Finally, the O-CU, which is splited into two components for the user plane (UP) and control plane (CP) functions, supports the higher layer protocols as SDAP, RRC, and PDCP. Furthermore, O-RAN specifies the O-Cloud platform, which hosts virtualized NFs (VNFs) from the \rev{O-gNB}.

To control and orchestrate the O-Cloud infrastructure and the \rev{O-gNB} functions, O-RAN introduces two Radio Intelligent Controllers (RICs): the non-real-time RIC (non-RT RIC) and the near-real-time RIC (near-RT) RIC. The Service Management and Orchestration (SMO) framework hosts the non-RT RIC, which enables control loops across large time-scales (i.e., seconds or minutes). Applications leveraging on the non-RT RIC control are called rApps. On the other hand, the near-RT RIC supports control loops on smaller time-scales (tenths of milliseconds) through applications called xApps.  

The O1 Interface is a logical connection between all O-RAN components and the SMO framework. The purpose of O1 interface is to ensure the operation and management i.e. fault, configuration, accounting, performance, and security (FCAP) of the O-RAN components. The components managed via O1 include the near-RT RIC, the O-CU, the O-DU. Moreover, The near-RT RIC uses the A1 interface to receive policies from the non-RT RIC, and E2 interface to collect near-real-time information from the O-RAN components and perform fine-grained radio resource management (RRM) policies over them. Finally the SMO performs O-Cloud management and orchestration via the O2 interface.

\section{Experimental Analysis}\label{sec:dissection}

We first investigate the root cause of the computing overhead when multiple vBS instances share a common GPP.  

\subsection{vRAN testbed}
To this end, we emulate a vRAN system with an off-the-shelf server and up to 10x software-defined radio (SDR) Ettus USRP B210 front-ends for both vBSs and the corresponding UEs which allows us to test up to 5 vBSs. The server provides an Intel i7-7700K CPU, with 4 physical cores and 8 virtual cores. The L1, L2, and L3 caches have 256 KiB, 1 MiB, and 8 MiB capacity, respectively. To implement a vBS, we use a \emph{full} 3GPP Rel.10-compliant stack from srsRAN~\cite{srslte} containerized with Docker, and we pair each vBS with one UE to generate downlink (DL) and uplink (UL) network load. Unless otherwise stated, the default bandwidth of each vBS is 10~MHz and we use $N=3$ physical cores in the experiments shown in this section. Using Docker's API, we developed a set of custom tools to dynamically orchestrate the vRAN system and configure different parameters related to the radio and the computing settings in run-time. 

\begin{figure}[t!]
 \centering
 \includegraphics[width=0.5\columnwidth]{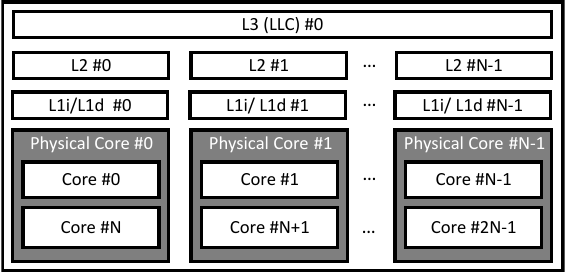}
 \vspace{-1mm}
 \caption{\small Hyper-threading \emph{vs.} no hyper-threading}
 \label{fig:diss:ht_vs_no_ht}
\vspace{-3mm}
\end{figure}

\subsection{Hyper-threading}

Previously, we described how modern processors employ SMT to optimize resource utilization within modern processor architectures. The impact of SMT on performance varies greatly depending on the application at hand. When two threads necessitate the processor's undivided attention, their execution can be hindered as they contend for processor access. However, if two threads engage in complementary tasks, with one requiring processor attention while the other focuses on reading and writing operations, SMT can yield significant cost-efficiency benefits by maximizing resource utilization. Fig. ~\ref{fig:diss:ht_vs_no_ht} shows the CPU utilization when we deploy different vBS at maximum traffic demand for uplink and downlink when using hyper-threading and when not using it. As we can see in the image, deactivating hyper-threading has a minimal performance improvement for less than 5 vBS. However, when deploying 5 vBSs in the system without hyper-threading, they cannot run with the maximum traffic demand as we have less computing capacity. Deactivating hyper-threading makes the platform more deterministic at the expense of having less computing capacity available. This is not surprising since the Linux CPU scheduler is aware of hyper-threading and leverages them through the scheduling domains \cite{bovet2005understanding}.

\vspace{-1mm}
\subsection{Network isolation}\label{sec:dissection:network}

Virtual networks incur substantial computing overhead. In the case of containers, the virtualization technology employed is a combination of network namespaces and virtual Ethernet pairs. With high data rates and small packet sizes, the number of operations that the host and the container must process consumes substantial CPU. This is a well-known problem reported in a plethora of literature~\cite{khalid2018iron, tootoonchian2018resq}. 

vBSs have (at least) two network interfaces: an interface with the backhaul, which connects vBSs to the mobile core (3GPP S1/Nn interfaces~\cite{3gpp136410}) and another one that connects to other vBSs (3GPP X2/Xn interface~\cite{3gpp136423}). For vRANs, network virtualization is not different than for traditional VNFs. Hence, we expect that common network isolation techniques, through \emph{network namespaces}, used in NFV behave similarly.

\begin{figure}[t!]
 \centering
\begin{subfigure}[t]{0.49\columnwidth}
\includegraphics[width=\columnwidth]{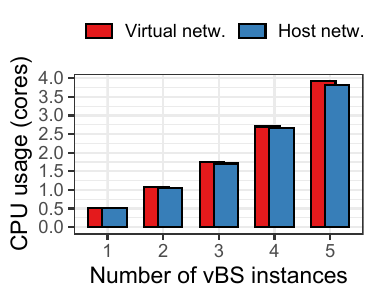}
\vspace{-7mm}
 \caption{\small{Virtual network \emph{vs.}\\ host network interface.}}
\label{fig:dissect:host_vs_virtual_netw}
\end{subfigure}
\hfill
\begin{subfigure}[t]{0.49\columnwidth}
\includegraphics[width=\columnwidth]{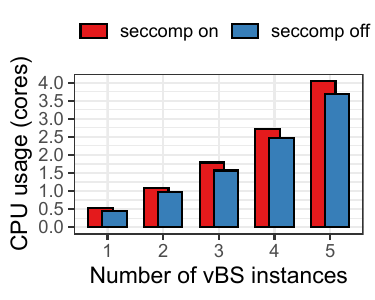}
\vspace{-7mm}
 \caption{\small{Seccomp enabled \emph{vs.}\\ Seccomp disabled.}}
\label{fig:dissect:seccomp}\end{subfigure}
 \caption{\small{95$^\text{th}$ percentile of aggregated per-core usage of a vRAN with different number of vBS instances.}}
 \vspace{-5mm}
\end{figure}

Fig.~\ref{fig:dissect:host_vs_virtual_netw} compares the mean CPU usage of scenarios with 1 to 5 vBS instances sharing the same physical network interface for backhauling. All vBS instances are homogeneous, with dedicated frequencies and we saturate their wireless capacity in both UL and DL directions. Moreover, in an attempt to reduce other potential sources of resource conflict, in this case we allocate each vBS on dedicated CPU cores.

We test two cases: ($i$) isolating the network stack of individual vBSs from the host using different network namespaces (``Virtual netw.''), and ($ii$) allowing all vBS to use host's networking without any namespace isolation (``Host netw.''). From the figure, we observe that the computing overhead of individual namespaces is negligible. The reason is that the aggregated network load generated by each BS is considerably smaller than the scenarios evaluated in the related literature~\cite{khalid2018iron, tootoonchian2018resq} (which handle over gigabit rates). Hence, network isolation cannot explain the computing toll showed in \S\ref{sec:intro}. 

\subsection{Secure computing filters}\label{sec:dissection:seccomp}

Docker containers (and others) use, by default in most modern GPPs, a security feature called Secure Computing (Seccomp) filters~\cite{seccomp},  Seccomp filters can control access to 300+ system calls (44 by default in Docker, which balances protection and compatibility). In the context of multi-tenant vRANs, this feature becomes of paramount importance to protect the underlying platform and mitigate potential attacks between potentially competing tenants. 

Though the overhead of seccomp filters is less studied in the literature, there exist some prior work that report a computing cost associated with seccomp filters that ranges from $<\!\!10\%$ (default seccomp profile in Docker) to almost $100\%$ (with an overprotective scheme)~\cite{9251949} with conventional applications. To complement that work, we now study the impact of seccomp filters in the context of vRANs. 

To this end, we deployed the same scenarios used in \S\ref{sec:dissection:network} (using virtual network interfaces) and measured the CPU usage without seccomp filters (``seccomp off'') and with the default seccomp profile in Docker (``seccomp on''). In line with \cite{9251949}, we observe a rough 1.4\% extra burden in CPU time for every vBS instance in the system, which adds up to 7\% total with 5 vBS instances. This is a non-negligible overhead, yet it does not fully explain the large toll observed in Fig.~\ref{fig:intro:vran}.

\subsection{Context switches}

The natural next step is to study the impact of context switches. Thread contention in shared CPUs may lead to an increased number of context switches and, consequently, increase the total consumption of CPU resources. 

\begin{figure}[t!]
 \centering
 \includegraphics[width=0.5\columnwidth]{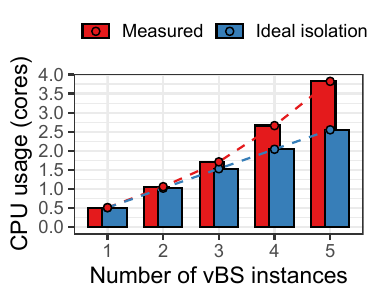}
 \vspace{-4mm}
 \caption{\small{95$^\text{th}$ percentile of aggregated per-core usage with different number of vBS instances and CPU pinning.}}
 \label{fig:cpu_iso:cpu}
\vspace{-4mm}
\end{figure}

To assess this, we repeat the same scenarios as before, and depict in Fig.~\ref{fig:cpu_iso:cpu}  the aggregated CPU usage for a variable number of vBS instances. Like before, we allocate dedicated CPU cores (CPU pinning) to individual vBS instances in an attempt to guarantee resource isolation. In the figure, we compare our empirical result with the expected outcome with ideal isolation. Though, as expected, the impact is considerable, it only accounts to 43\% of the overhead observed in Fig.~\ref{fig:intro:vran}. 

\begin{figure}[t!]
     \centering
     \begin{subfigure}[t]{0.49\columnwidth}
         \centering
         \includegraphics[width=0.93\columnwidth]{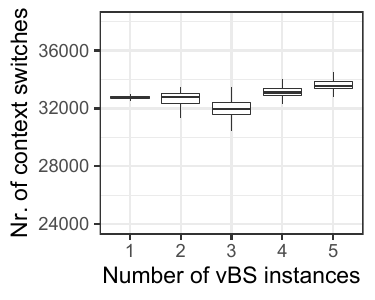}
            \vspace{-2mm}
         \caption{\small With pinning}
         \label{fig:cpu_iso:pin}
     \end{subfigure}     
     \hfill
     \begin{subfigure}[t]{0.49\columnwidth}
         \centering
         \includegraphics[width=0.93\columnwidth]{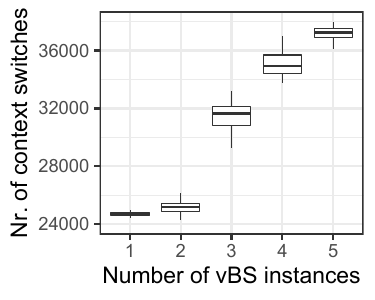}
            \vspace{-2mm}
         \caption{\small Without pinning}
         \label{fig:cpu_iso:nopin}
     \end{subfigure}
     \vspace{-2mm}
        \caption{\small Context switches per ms experienced by one vBS.}
        \vspace{-5mm}
        \label{fig:cpu_iso}
\end{figure}

To gain more insights, Fig.~\ref{fig:cpu_iso} compares the ratio of context switches experienced by an individual vBS in two different settings: ($i$) when each vBS is pinned to an individual CPU (as in \S\ref{sec:dissection:network}), in Fig.~\ref{fig:cpu_iso:pin}; and ($ii$) when the default CPU scheduler is free to allocate threads within the shared CPU pool (as in the experiment of \S\ref{sec:intro}), in Fig.~\ref{fig:cpu_iso:nopin}.

From Fig.~\ref{fig:cpu_iso:pin}, we observe that the ratio of context switches remains very similar irrespective of the number of vBSs deployed. In this case, all the CPU contention is caused by the threads that belong to the sampled vBS. Since these are homogeneous vBSs (which implement the same amount of threads), and each of them is pinned to a dedicated CPU, the amount of contention in individual CPUs is independent of the number of vBSs deployed. 

We observe a different behavior in Fig.~\ref{fig:cpu_iso:nopin}. In this case, the threads of all the vBSs compete for the same pool of CPUs. Surprisingly, when the number of vBS instances deployed in the platform is 1 or 2, the ratio of context switches is \emph{smaller} than that when vBSs use dedicated CPUs. The reason is that the number of instances (1 or 2) is relatively smaller than the number of CPUs in the pool (6 virtual cores with $N=3$). Hence, individual threads often find less contention than in the setting used for Fig.~\ref{fig:cpu_iso:pin} because, there, individual CPUs are dedicated to individual vBS instances but they are shared between the threads implementing the vBS (intra-vBS contention). Conversely, when the number of instances is close to the number of CPUs in the pool (4 and 5), inter-vBS thread contention dominates and the ratio of context switches noticeably overpasses that when CPUs are dedicated to individual vBSs. Interestingly, when we deploy 3 vBS instances, intra-vBS and inter-vBS thread contention balance out and the ratio of context switches is similar to the case when vBSs are pinned to dedicated CPUs.

With 5 vBS instances, we measure a rough 8\% increase in context switches when there is no pinning with respect to using CPU pinning. Moreover, when just one vBS is deployed, there is a 24\% decrease in the number of context switches that does not translate into a reduction in overall CPU time usage.
Consequently, context switching cannot explain the aforementioned 43\% increase in the overall CPU consumption observed in Fig.~\ref{fig:intro:vran} with respect to  Fig.~\ref{fig:cpu_iso:cpu}, which lead us to the next subsection. 

\subsection{Cache memory isolation}

Cache memory is a very relevant resource that is often overlooked. Although Docker provides efficient mechanisms to partition and isolate different types of resources, it does not provide features to partition cache memory resources effectively. However, cache-intensive applications sharing memory resources tend to evict each other’s cache values, which increase the number of cache misses~\cite{intel2015improving}. As explained in \S\ref{sec:background}, cache misses cost additional CPU cycles. If data is not available in a low-level cache, a core executing a thread will trigger an interrupt signal that halts its execution until the corresponding value is finally retrieved from some higher-level memory resource. This cost in CPU cycles differ across technologies. However, we can infer its order of magnitude by observing the  latency required to access different types of memory. As a reference, Table~\ref{tab:cache_cycles} shows the latency to access different cache levels in an Intel Skylake architecture. 

\begin{figure}[t!]
\vspace{-2mm}
\minipage{0.47\columnwidth}
\centering
 \includegraphics[width=\columnwidth]{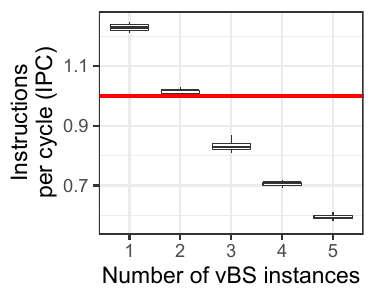}
 \vspace{-4mm}
 \caption{\small Instructions per cycle (IPC) of a vBS}
 \label{fig:diss:ipc}
\endminipage
\hfill
\minipage{0.47\columnwidth}
 \includegraphics[width=\columnwidth]{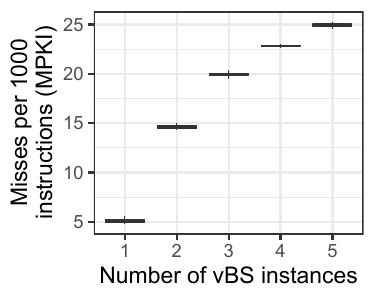}
 \vspace{-7mm}
 \caption{\small Misses per 1000 instructions (MPKI) of a vBS}
 \label{fig:diss:mpki}
\endminipage
\vspace{-7mm}
\end{figure}

To study the impact of cache contention in vRANs, we used the tool \texttt{perf} to measure the ratio of cache misses, CPU cycles and instructions required by one vBS in a system with 1-to-5 vBS instances. These measurements are summarized in Figs.~\ref{fig:diss:ipc} and \ref{fig:diss:mpki}, which show, respectively, the instructions executed per cycle (IPC), and  the number of cache misses per 1000 instructions (MPKI).
Both metrics show high correlation.

\begin{table}[t!]
\footnotesize
\centering
\begin{tabular}{c|c|c|}
Memory type & Access latency\cite{patterson2003modern, jacob2010memory, drepper2007every} \\\hline
L1 cache & 4-6 cycles \\
L2 cache & 14 cycles\\
L3 cache & 50-70 cycles \\
RAM & $\sim$ 120 - 600 cycles \\
\end{tabular}
\caption{\small Access and cache miss latency}
\label{tab:cache_cycles}
\vspace{-3mm}
\end{table}

Fig.~\ref{fig:diss:ipc} evinces that an increasing number of vBS instances has a huge impact on computing efficiency. The red line indicates a boundary point of operation where the system process 1 instruction per cycle~\cite{gregg2014systems}. On the one hand, when IPC $ > 1$, the application is instruction-bounded, i.e., only improving the efficiency of the software code can improve the IPC performance further. On the other hand, when IPC $ < 1$, the application is likely bounded by a bottleneck when accessing resources other than CPU, such as memory. In the case of Fig.~\ref{fig:diss:ipc} the latter occurs for a number of vBS instances larger than 2. Such a bottleneck is remarkable, allowing only 0.6 instructions per cycle when 5 vBSs are instantiated. 

Conversely, Fig.~\ref{fig:diss:mpki} shows a dramatic growth of cache misses per instruction, a 500\% increase with 5 vBSs with respect to 1. This, and the strong correlation between cache misses and IPC dynamics, lead us to infer that cache memory is the bottleneck in our vRAN system and, ultimately, the root cause of the anomalous CPU behavior shown in Fig.~\ref{fig:intro:vran}. 

There exist mechanisms that can alleviate the impact of cache contention on CPU consumption. Perhaps the most effective approach is Intel Cache Allocation Technology (CAT)~\cite{intel2015improving}, which allows us to partition cache memory resources among different applications. Unfortunately, standard virtualization technologies based on cgroups (such as Docker containers) do not support such a mechanism natively. Hence, we need to find alternative strategies that allocate CPU resources to vBS instances \emph{considering the impact of \rev{noisy neighbours problem}}, which motivates our next section.

\vspace{-1mm}
\section{\name{} design}\label{sec:design}

In this section, we first formalize our problem and then we describe our proposed solution, named \name{}. \name{} aims to minimize the operating cost of the vRAN infrastructure (based on CPU usage). To this end, \name{} \emph{learns} the relationship between vBS instances, which incur resource contention in the computing platform, and network performance to optimize the allocation of computing resources in the system.

 \begin{figure}[t!]
  \centering
 \minipage{0.35\columnwidth}
  \includegraphics[width=\columnwidth]{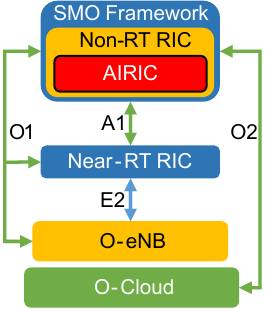}
 \caption{\small \name{} within O-RAN}
  \label{fig:\name{}:system}
 \endminipage{}
 \hfill
 \minipage{0.47\columnwidth}
 \includegraphics[width=\columnwidth]{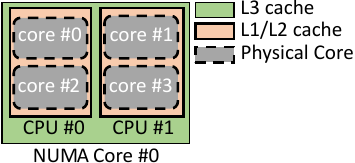}
 \caption{\small Toy GPP}
 \vspace{-8mm}
  \label{fig:airic:toy_scn}
 \endminipage{}
 \vspace{-5mm}
 \end{figure}

\vspace{-2mm}

\subsection{The problem}

The computing requirements of a vRAN system are hard to quantify dynamically. To begin with, the amount of CPU resources required by a single vBS instance depends on the network traffic demand on both DL and UL directions, the signal-to-noise ratio (SNR) of each wireless link and the associated Modulation Coding Scheme (MCS) used for communication, in a non-trivial manner~\cite{vrain, vrain2, nuberu}. Moreover, estimating the actual requirements for a set of vBS instances sharing a platform is even more challenging because the overhead introduced by computing resource contention (\rev{noisy neighbours problem}) depends on the computing cores used to process each vBS workload, the amount of isolation across vBS instances, and the maximum computing capacity available. 

On the one hand, over-dimensioning the allocation of computing resources incurs high infrastructure costs as many computing cores might not be needed when running a small number of vBS instances or when the aggregated load is low, and the electricity bill associated with unneeded active cores can be substantial. 
On the other hand, pooling a reduced number of cores across many instances (i.e., forcing vBSs to share) may lead to throughput loss because heavy resource contention leads to severe computing overheads. As we demonstrated in \S\ref{sec:intro}, a shortage of computing resources (due to the influence of \rev{the noisy neighbors problem}) may cause that the users associated with vBSs in the system lose synchronization, induce a high number of radio link errors, and cause very high end-to-end latency and jitter. 

Moreover, though pinning vBS workloads to specific CPU cores provides better isolation and performance determinism, as shown before, it requires activating a larger pool of CPU cores, which incurs higher energy costs. Hence, our approach is to let all the vBS instances fairly share a pool of CPU cores, using a standard scheduler, and determine dynamically the  smallest set of active CPU cores in the pool at every time step to minimize energy costs. The key novelty in our approach is that we do so in a reliably manner, accounting for the costs of sharing, as dissected earlier. As we show later in \S\ref{sec:evaluation}, ignoring such cost has dramatic consequences on network performance.

\vspace{-3mm}

\subsection{System model} \label{sec:system_model}

We consider an O-RAN cloud computing platform (O-Cloud) providing computing resources for multiple vBS instances deployed therein, i.e., each vBS instance shares the same pool of computing resources. We also consider an agent in charge of ($i$) observing the \emph{context} associated with each vBS, and ($ii$) devising which computing cores need to be active in the pool to serve the demand of each vBS, which process uplink and downlink traffic. As shown in Fig.~\ref{fig:\name{}:system}, following O-RAN's specification, our agent is hosted by the system's Service Management and Orchestration (SMO), and takes decisions in discrete time intervals $t \in \mathbb{N}$, which we call \textit{decision intervals} and are in the range of several seconds to minutes following O-RAN's specification for the Non-Real-Time RAN Intelligent Controller (Non-RT RIC). 

Our agent employs an O-RAN-compliant monitoring system that gathers metrics from the various O-RAN components (such as O-RU, O-DU, and O-CU) and measurements from the O-Cloud platform (i.e. infrastructure metrics). The near-RT RIC uses the E2 interfaces to periodically receive different radio metrics from the components deployed in the O-Cloud platform~\cite{oranwg3e2}. Afterward, the near-RT RIC passes the data using the O1 interface to the non-RT RIC. On the other hand, to gather metrics from the O-Cloud platform, the agent sets up performance management (PM) jobs that collect different infrastructure metrics (i.e. computing usage, energy consumption) using the O2 interface~\cite{oranwg6o2}. Finally, to enforce the different computing policies that our agent computes, it uses the O2 interface to pass those policies to the O-Cloud platform. Fig.~\ref{fig:\name{}:system}, depicts how our agent integrates into the ORAN architecture

Given the hard-to-model nature of the \rev{noisy neighbour problem}, we advocate for reinforcement learning (RL) to design our agent. In this way, the agent observes the context and takes an action at the beginning of each decision interval, and then receives a reward at the end of the decision interval. The learning agent stores $3$-tuple samples comprised of the context, actions, and the associated rewards at every interval, and uses these experiences to learn and improve the obtained rewards over time. Note that while the admission control problem is out of the scope of this paper, we do support a number of active vBS instances that may vary over time. To the best of our knowledge, this is the first solution that optimally allocates computing resources in a vRAN system accounting for the overhead of \rev{the noisy neighbours problem} and a dynamically changing number of vBS instances in the system. 

\begin{figure}[t!]
 \centering
 \includegraphics[width=\linewidth]{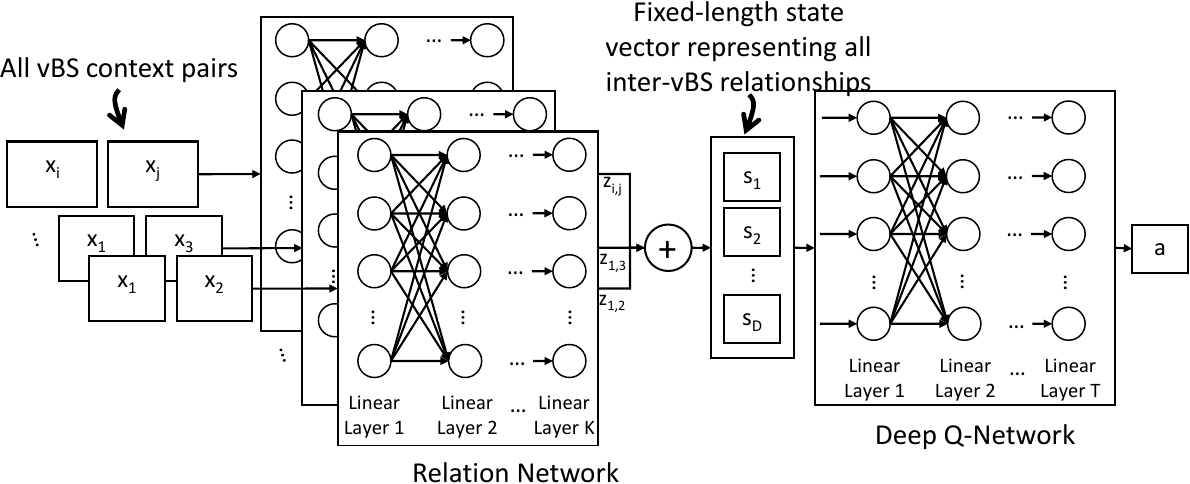}
 \caption{\name{} Machine Learning Architecture}
 \vspace{-4mm}
 \label{fig:arch}
\end{figure}

\subsection{Optimization framework}

A variable number of vBS instances imply that the dimensionality of the context information also varies over time. This is particularly challenging to support with standard RL solutions. To address this, we augment a classical Deep Q-Network (DQN) approach~\cite{mnih2013playing} with a Relation Network (RN) mechanism~\cite{raposo2017discovering} as shown in Fig.~\ref{fig:arch}. 

The basic idea of an RL agent is to learn an optimal policy $\pi$ by interacting with an environment $\mathcal{E}$ in discrete time intervals $t \in \{1, 2, \dots, T\}$. Every interval, an agent observes a state (or context) $\vv{s}^{(t)}$, selects an action $a^{(t)}$ and receives a reward $r^{(t)}$ at the end of the time step. A policy $\pi$ is a distribution of actions over the different states, which captures the \emph{goodness} of the state-action pair $(\vv{s}^{(t)}, a^{(t)})$. Once the reward $r^{(t)}$ is measured,  the system transitions to state $\vv{s}^{(t+1)}$. After $T$ intervals, $\mathcal{E}$ reaches its terminal state and the agent refines its policy $\pi$ using past observations $\{\{\vv{s}^{(1)}, a^{(1)}, r^{(1)}\}, \dots, \{\vv{s}^{(T-1)}, a^{(T-1)}, r^{(T-1)}\}\}$. The goal is to maximize the total discounted reward $R^{(t)} := r^{(t)} + \sum_{t' = t+1}^T \gamma^{t'} r^{(t')}$. 

Most RLs approximate value functions that estimate the importance of actions given a state $\vv{s}$. One of the those value functions is $Q^*(\vv{s}, a) := \max_{\pi}\EX[R^{(t)} | \vv{s}^{(t)} = \vv{s}, a^{(t)} = a]$, which represents the maximum expected return given an action-state pair under the policy $\pi$. The optimal $Q^*$-value function follows the Bellman Optimality Equation, which provides $Q^*(\vv{s}^{(t)}, a^{(t)})$ in terms of $Q^*(\vv{s}^{(t+1)}, a^{(t+1)})$: 
\begin{align}
Q^*(\vv{s}, a) &= \EX\Big[r^{(t)} + \nonumber\\ 
&\gamma\max_{a^{(t + 1)}} Q^*(\vv{s}^{(t+1)}, a^{(t+1)})  | \vv{s}^{(t)} = \vv{s}, a^{(t)} = a\Big]  \nonumber
\end{align}
Using the Bellman Optimality Equation, we can  find $Q^*(\vv{s}, a)$ iteratively~\cite{sutton2018reinforcement}. In this paper, we have used neural networks to approximate the optimal $Q^*(\vv{s}, a)$, which is called \rev{Deep Q-Network} (DQN)~\cite{mnih2013playing}. 
In particular, given the large timescale of the Non-RT RIC, the action taken at one interval $a^{(t)}$ has little impact on the next state $\vv{s}^{(t+1)}$ and therefore it is enough to maximize instantaneous reward. Hence, to expedite convergence, we simplify our RL setting into a contextual bandit problem by setting $\gamma=0$ and $T=1$.

We next describe our design for the learning agent's context (states), actions, and reward function.

\subsubsection{Context}
In line with the related literature~\cite{vrain, vrain2, baysesian, edgebol}, we use the next metrics to describe the state:
\begin{itemize}
    \item \textbf{Chanel quality}: We use the mean UL SNR observed by each vBS in the last interval, which allows our agent to infer their UL wireless capacity, and the mean DL channel quality indicator (CQI) to do the same for the DL. 
    \item \textbf{Network demand}: The network demand of a vBS is the amount of UE buffered data for both UL and DL during the last decision interval. 
\end{itemize}
We represent DL and UL channel quality for a vBS instance $i$ observed in interval $t$ as $\sigma_{\text{DL},i}^{(t)}$ and $\sigma_{\text{UL},i}^{(t)}$. Furthermore, we let $d_{\text{DL},i}^{(t)}$ and $d_{\text{UL},i}^{(t)}$ denote its DL and UL network demand, respectively. We also assume a known mapping between channel quality and MCS: $g_{\text{DL}}(\sigma_{\text{DL},i})$ for DL, $g_{\text{UL}}(\sigma_{\text{UL},i})$ for UL, which is a mild assumption. Because the channel quality bounds the highest MCS, we can estimate the mean number of radio Resource Blocks (RBs) that each vBS can use in both directions given a mean MCS and network demand. This can be estimated using the 3GPP specifications~\cite{3gpp136213}. In this way, we can state the demand for radio resources (RBs) rather than relying only on the past utilization of Radio Blocks, which may differ. Consequently, we denote the number of RBs used for DL and UL for vBS $i$ as $p_i^{\text{DL}}$ and $p_i^{\text{UL}}$, respectively. Using the number of RBs and network demand, we define the context of vBS $i$ as

$$\vv{x}_i^{(t)} := (p_{\text{DL},i}^{(t)}, d_{\text{DL},i}^{(t)}, p_{\text{UL},i}^{(t)}, d_{\text{UL},i}^{(t)}) $$
The design of $\vv{x}_i$ is motivated by the convenience of expressive features and minimal dimensionality and follows the state of the art~\cite{vrain, vrain2, baysesian, edgebol}. The challenge now is to encode the context information  $\{\vv{x}_i\}$ for all vBS instances $i$ in a state vector $\vv{s}$ with \emph{fixed} dimensionality $D$, which is required by the DQN model, in scenarios with a variable number of vBS instances over time. As shown in Fig.~\ref{fig:arch}, we address this with a Relation Network (RN)~\cite{mnih2013playing}. 

\subsubsection{Relation Network}

As the number of vBSs that \name{} has to allocate CPU resources for in a particular time interval might be different than in past intervals, the context length changes depending on the number of vBS instances. Rather than building other agents for each of the different numbers of vBS cases or padding the various possible contexts to match a fixed context length, we opted to solve the problem using a Relation network. A RN can encode the relationship between the context associated to all vBS instances into a fixed-length state vector $\vv{s}$.  
To this end, the RN operates along all possible pairs of objects (context of vBS instances) to capture such hidden relations with a multi-layered perceptron (MLP) model. Assuming a maximum number of vBS instances supported in the system equal to $M$, then we have the following possible pairs of context vectors:

$$\mathcal{X}:=\{ (\vv{x}_1,\vv{x}_2), (\vv{x}_1, \vv{x}_3), ... (\vv{x}_{M-1}, \vv{x}_{M}) \}$$

Since the maximum amount of vBS instances at any given moment is bounded, then $|\mathcal{X}|$ is also bounded and fixed over time. The RN ingests sequentially each pair $(\vv{x}_i,\vv{x}_j) \in \mathcal{X}$ of possible unpermuted context combinations, and generates an output vector $\vv{z}_{i,j}$ with cardinality $D$. 
Once all ${N\choose 2}$ permutation vectors $\vv{z}_{i,j}$ are computed by the RN, which is done sequentially, we create an encoded state vector $\vv{s}$ by aggregating all output vectors, i.e., $\vv{s} = \sum_{i,j} \vv{z}_{i,j}$
In this way, we force order permutation invariance, which is a critical requirement of our problem, i.e., as the RN learns about different latent relations across vBS instances (objects), these learned relations remain invariant regardless the order of the input pair relations. Importantly, our RN not only helps to support variable number of vBS instances over time, it also provides the DQN model with state information that represents better the relations between them, \emph{which is very helpful to capture the impact of the \rev{noisy neighbours problem} in a state dimension-fixed representation}. To this end, we train the RN network jointly with the DQN model as we explain later. 

\subsubsection{Actions}

Given state $\vv{s}^{(t)}$, our agent shall \emph{activate} the appropriate set of CPU cores, described with an \emph{activation vector} $\vv{v}$ wherein each element corresponds to the CPU core index that shall be activated. Then, all the vBS instances will fairly share the pool of CPU cores in $\vv{v}$. By avoiding pinning vBS workloads into specific cores, we aim at maximizing resource multiplexing and, consequently, at reducing the overall usage of computing resources. To ensure quick convergence, we need to preserve a low action space dimensionality. To address this we resolve our action into two steps. In step 1, our RL agent decides the \emph{total} number of CPU cores that shall be activated to \emph{guarantee service.} Thus, the set of actions $A$ is $A=\{1, 2, \dots , 2N\}$, where $N$ is the total number of physical cores available. Then, in step 2, we implement a \emph{deterministic} rule $\rho (a)$ to \emph{minimize infrastructure cost}. That is, $\rho \colon A \to \mathcal{V}_a, a \mapsto \vv{v}$,
where $\mathcal{V}_a$ is a set containing all possible activation vectors such that $a=|\vv{v}|$. Because $\rho$ is a pre-determined rule to minimize cost, the agent can learn its policy $\pi$ to guarantee service given $\rho$ as part of the environment $\mathcal{E}$.

\begin{figure}[t!]
 \centering
 \includegraphics[width=\linewidth]{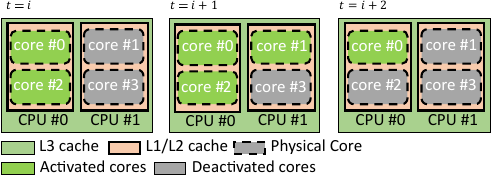}
 \caption{\name{} actions timeline}
 \vspace{-4mm}
 \label{fig:airic_time}
\end{figure}

See, e.g., the GPP of Fig.~\ref{fig:airic:toy_scn} with $N=2$. If $a=1$ then $\mathcal{V}_{a=1}\!=\!\{(0), (1), (2), (3)\}$ all the activation vectors in $\mathcal{V}_{a=1}$ are equivalent and any $\vv{v}\in\mathcal{V}_{a=1}$ could be chosen trivially. However, this is not necessarily the case for other actions $a$ because, as we explained before, modern processors leverage multi-processing CPUs, being two virtual cores for each physical CPU the most common case. For instance, for $a=2$ (and the same GPP with $N=2$), the set of possible activation vectors is $\mathcal{V}_{a=2}=\{(0, 2), (1, 3), (0, 1), (0, 3), (1, 2), (1, 3)\}$. Though many of the vectors in $\mathcal{V}_{a=2}$ are equivalent, others are not. Subset $\hat{\mathcal{V}}_{1,a=2}=\{(0,2), (1,3)\}\subset\mathcal{V}_{a=2}$ contains equivalent activation vectors; and so are the activation vectors in $\hat{\mathcal{V}}_{2,a=2} =\{(2, 3), (0, 1), (0, 3), (1, 2)\}\subset\mathcal{V}_{a=2}$. But any $\vv{v}_1 \in \hat{\mathcal{V}}_{1,a=2}$ and any $\vv{v}_2 \in \hat{\mathcal{V}}_{2,a=2}$ are \emph{not} equivalent. On the one hand, any $\vv{v}_1 \in \hat{\mathcal{V}}_{1,a=2}$ incurs more cache contention than any $\vv{v}_2 \in \hat{\mathcal{V}}_{2,a=2}$ because all the cores in $\vv{v}_1$ share the same physical CPU (see Fig.~\ref{fig:airic:toy_scn}). On the other hand, any $\vv{v}_2 \in \hat{\mathcal{V}}_{2,a=2}$ is more costly than any $\vv{v}_1 \in \hat{\mathcal{V}}_{1,a=2}$ because $\vv{v}_1$ allows turning off more physical CPUs, e.g., if $\vv{v}_1=(0,2)$ CPU 1 can be turned off (see Fig.~\ref{fig:airic:toy_scn}). Fig.~\ref{fig:airic_time} illustrates an example of the operation of our algorithm during three time steps. Importantly, give a pool of activated CPU cores, all vBS instances will fairly use those cores using a standard scheduler.

In the assumption that, given any static mapping $\rho$, policy $\pi$ will provide an appropriate cardinality for the activation vector to guarantee network service ($a=|\vv{v}|$), we just need to design $\rho$ aiming to minimize the amount of infrastructure (physical CPUs) that has to be activated given $a$. Consequently, we propose the following simple rule. Let $k(\vv{v}) \in \{1, 2, \dots, N\}$ denote the number of physical CPUs that contain at least one virtual core activated in $\vv{v}$. Then, given a set $\mathcal{V}_a$ with all possible activation vectors for action $a$, we define the \emph{ordered} superset $\mathcal{W}_a:= \langle \hat{\mathcal{V}}_{1,a}, \dots, \hat{\mathcal{V}}_{N,a} \rangle$, where $\mathcal{V}_{i,a} = \{ \vv{v} | k(\vv{v})=i, \vv{v} \in \mathcal{V}_a \}$. In the example above, with $a=2$ and $N=2$, $\mathcal{W}_a = \{ \hat{\mathcal{V}}_{1,a=2}, \hat{\mathcal{V}}_{2,a=2}\}$. Note that $\hat{\mathcal{V}}_{i,a}=\emptyset$  for some $i$. For instance, in our toy example with $N=2$, $\hat{\mathcal{V}}_{1,a=3}=\emptyset$ for $a=3$. Hence, we let $\rho(a) = \vv{v}  \in \hat{\mathcal{V}}_{m,a}$ such that $m := \argmin_i\{i\,\,| \hat{\mathcal{V}}_{i,a} \neq \emptyset\}$.

\subsubsection{Reward}

Our goal is to meet the traffic demand of all the vBS deployed in the system over time with minimum physical infrastructure (to save costs by turning off CPUs). Assuming a pool with $N$ physical CPUs and $2N$ virtual cores, where cores $j$ and $j+N$ belong to the same physical CPU $\forall j < N$, we let $z(j) \in {\{0,\dots,2N-1\}}$ denote the \emph{sibling} virtual core $i$ given input virtual core $j$. A sibling core is that that uses the same physical CPU. For instance, in the toy GPP of Fig.~\ref{fig:airic:toy_scn}, with $N=2$ physical CPUs and $4$ cores, $z(0) = 2$ and $z(2) = 0$. 

Following the related literature~\cite{jaiantilal2010modeling, shahid2019energy}, we codify the cost associated to an activation vector $\vv{v}$ using a linear model. Let us first denote $c_j^{(t)}\in [0,1]$, as the relative usage of computing core $j$ during interval $t$. If $j\, \notin \, \vv{v}^{(t)}$, then $c_j^{(t)} = 0$; otherwise, $c_j^{(t)}$ is empirically measured. Then, we let $E_j^{(t)}$ model the (energy-related) cost associated with computing core $j \in \{0, 1,..., 2N-1\}$ as follows:

\begin{align}
 E_j^{(t)} := \begin{cases} 
      \alpha_1 + \beta \cdot c_j^{(t)} &  \text{if } c_j^{(t)} > 0 \\
      \alpha_2 & \text{if } c_j^{(t)} = 0 \text{ and } c_{z(j)}^{(t)} > 0 \\
      \alpha_3 & \text{if } c_j^{(t)} = 0 \text{ and } c_{z(j)}^{(t)} = 0
   \end{cases}
\end{align}
where $\alpha_1 > \alpha_2 > \alpha_3$. Intuitively, $\alpha_i$ models the bias cost of a core, which is different depending on the activation state of core $j$ and its sibling. We choose $\alpha_i$ and $\beta$ so that $0\le E_j\le 1$.

We now let $\tau_{DL,i}^{(t)}$ and $\tau_{UL,i}^{(t)}$ denote the DL/UL throughput experienced by vBS $i$ during interval $t$, and then formalize our reward function as:
\begin{align}
r^{(t)} \!:=\! 
\begin{cases}
     -1, & \text{if } \tau_{DL,i}^{(t)} < d_{DL,i}^{(t)}\, \text{for any $i$}\\
     -1, & \text{if } \tau_{UL,i}^{(t)} < d_{UL,i}^{(t)}\, \text{for any $i$}\\
     \frac{1}{2N}\sum_{j=0}^{2N-1}{ -E_j }, &  \text{otherwise}
\end{cases}
\end{align}

\subsubsection{Training}

As explained above, the goal is to train a policy to approximate an optimal action-value function $Q^*$. Our policy $\pi$ is implemented by the structure of RN+DQN introduced above and, hence, we shall optimize the weights $\vv{\Theta} := (\vv{\theta_1}, \vv{\theta_2})$ of the combined neural networks to estimate the Q-value function $Q(s, a; \theta) \approx Q^*(s, a)$. To this end, we use a Smooth L1-loss function~\cite{girshick2015fast}. 

\begin{align}
    L^{(t)} (\Theta_i) &:=
    \begin{cases}
         \frac{1}{2} \frac{x^2}{1} & \text{if } |x| < 1 \\
        |x| - \frac{1}{2}\cdot 1 & \text{otherwise}
    \end{cases}
\end{align} 
where $x = \EX_{(s, a, r, s') \sim \rho}[(y_i - Q(s, a; \Theta_i))]$ and $y_i = r + \gamma \max_{a^{(t+1)}} Q(s', a^{(t+1)}; \Theta_{i-1})$. 
$\rho$ is a replay buffer from where we sample  $(s, a, r, s')$, $y_i$ is the temporal difference target, and $y_i - Q$ is the temporal difference error. We use a target network to stabilize the training process, that is, the learning agent uses a different target network with fixed weights that are used to compute the loss function used in turn to train the primary $Q$-network. It is crucial to stress that the target network's parameters are periodically synchronized with those of the primary Q-network rather than being trained. The primary Q-network is trained using the target network's Q values in an effort to increase the training's stability. Finally, we use a standard $\epsilon$-greedy approach for exploration.

\section{Performance Evaluation}\label{sec:evaluation}

\begin{figure}[t!]
 \centering
 \includegraphics[width=\linewidth]{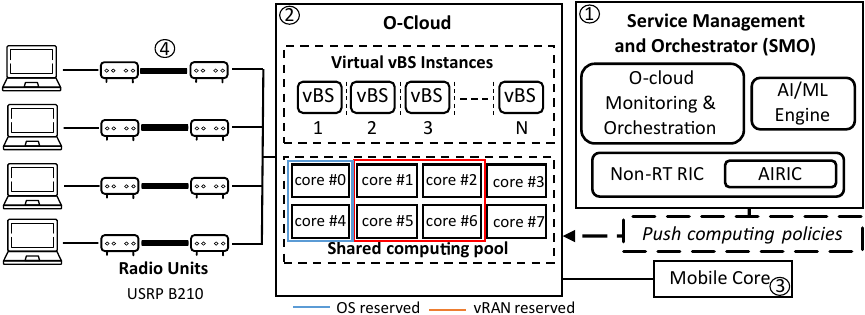}
 \caption{Conceptual design of the evaluation testbed}
 \vspace{-4mm}
 \label{fig:concept_testbed}
\end{figure}

We have built an O-RAN-compliant experimental testbed to evaluate \name{}. The testbed comprises different hosts, which contain the components of an O-RAN deployment and the ones to provide network connectivity to different connected UEs. Fig.~\ref{fig:concept_testbed} depicts conceptually the testbed that we have built. First, this testbed has a host, which deploys the SMO and contains the non-RT RIC where we deploy \name{} (\circled{1}). Second, it has a separate host that hosts the O-Cloud platform where different O-eNB instances can be deployed and also comprises the near-RT RIC \circled{2}. To implement the orchestration and management functions of the O-Cloud platform provided by the SMO, we have opted to implement the O-eNBs deployed in the O-Cloud platform, containerizing srsRAN using Docker. Thus, we use Docker API capabilities to orchestrate and manage containers to implement a minimal O2 interface. In addition, we used a metrics agent as Telegraf to implement the performance monitoring jobs, which allowed us to gather metrics from the O-Cloud platform. Rather than using a commercial orchestrator such as Kubernetes or Docker swarm, we implemented our minimal orchestrator for performance and flexibility. Moreover, we have also implemented minimal O1 and E2 interfaces to allocate resources on the vBSs deployed. Our testbed also includes a host, which contains the EPC to provide connectivity to the different UEs attached to each vBS \circled{3}. As the vBSs are containerized using Docker we have isolated the networking from each one another. 

The O-Cloud host comprises an Intel i7-7700K GPP with 4 physical CPUs. We use Ubuntu $20.04.5$ LTS with kernel $5.13.19$. We reserve 1 physical CPU (2 virtual cores) for the OS and custom scripts to manage the experiments, interact with Docker API, and collect data, i.e., we emulate a small GPP vRAN platform with $N=2$ physical CPUs and 4 virtual cores (as in Fig.~\ref{fig:airic:toy_scn}). The testbed also integrates 4 USRP SDR boards to support up to 4 vBS (and the corresponding UEs to generate network load) \circled{4}. To generate uplink and downlink flows, we use mgen\footnote{https://github.com/USNavalResearchLaboratory/mgen} to initiate a flow from/to the UE to/from the EPC. Given the constrained computing capacity of our testbed, we set the bandwidth of each vBS to 10~MHz. We have generated 60k context-action-reward data samples, evenly split for scenarios with 2, 3 and 4 vBS instances operating concurrently. We shuffled and split the dataset into a training and a testing set of 40k and 20k samples, respectively.\footnote{Our dataset will be publicly available upon publication.}

We have implemented \name{} using PyTorch\footnote{\url{http://www.pytorch.org}}. On the one hand, the RN has one hidden layer and the same number of neurons than the output layer, 128. On the other hand, the DQN has one hidden layer with 256 neurons. The initial parameters of the neural networks are initialized from an uniform distribution. We also use the ReLu activation function, and a normalization layer~\cite{ba2016layer} in between hidden layers. For the $\epsilon$-greedy mechanism, we use a decay factor equal to $60\%$ of the size of the training set.
We also use a replay buffer with 20k samples and batches of 128 samples. Finally, we used Adam \cite{kingma2014adam} as our optimizer. These implementation choices are intended to stabilize training based on ~\cite{ba2016layer, thimm1997high}. 

\subsection{Convergence Evaluation}

\begin{figure}[t!]
     \centering
     \begin{subfigure}[b]{0.49\columnwidth}
         \centering
         \includegraphics[width=\columnwidth]{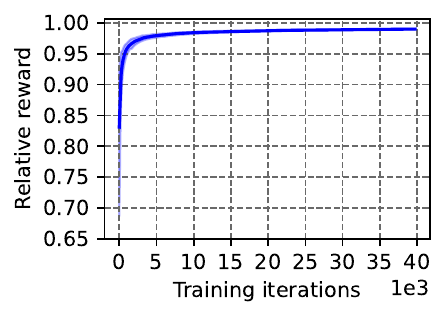}
          \vspace*{-4mm}
         \caption{\small Randomized contexts}
         \vspace*{-3mm}
         \label{fig:eval:convergence_normal}
     \end{subfigure}
     \hfill
     \begin{subfigure}[b]{0.49\columnwidth}
         \centering
         \includegraphics[width=\columnwidth]{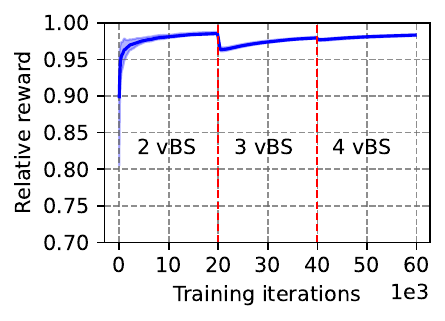}
              \vspace*{-4mm}
         \caption{\small Sequential number of vBSs}
         \vspace*{-3mm}
         \label{fig:eval:convergence_order}
     \end{subfigure}
     \vspace{-2mm}
        \caption{AIRIC Convergence Evaluation}
        \label{fig:eval:convergence}
    \vspace{-2mm}
\end{figure}

We start evaluating convergence. Fig.~\ref{fig:eval:convergence} shows the normalized reward of \name{} over training iterations. The UL/DL load and SNR generated in both plots are chosen uniformly at random. However, while the number of vBS instances is also random (between 2 and 4) in Fig.~\ref{fig:eval:convergence_normal}, they arrive sequentially in Fig.~\ref{fig:eval:convergence_order}. In the former case,  the reward  converges to 0.95 in less than 5k iterations. In the latter case, there are expected bumps when new vBSs arrive but these are small, within  $5\%$. Hence, we conclude that the RN in \name{} learns correctly the relationship across vBSs and how to use its experience to quickly reach close-to-optimal performance.

\begin{figure}[t!]
 \centering
 \includegraphics[width=0.5\linewidth]{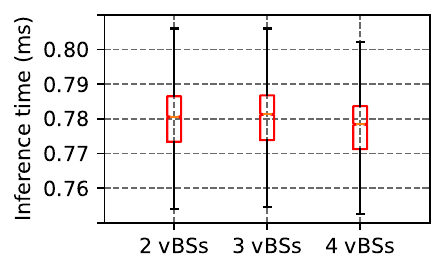}
 \caption{AIRIC's inference time}
 \vspace{-6mm}
 \label{fig:inf_time}
\end{figure}

\subsection{Inference time}
In order to assess whether \name{} is suitable for running in a non-RT RIC controller, we measured the inference time of our approach for the different number of vBS cases. The results, depicted in Fig.~\ref{fig:inf_time}, shows inference times lower than 1 millisecond (ms) for all cases, which is well below the control-loop cycle of a RIC controller and validates \name{} to operate therein appropriately.

\begin{figure}[t!]
\centering
 \includegraphics[width=\columnwidth]{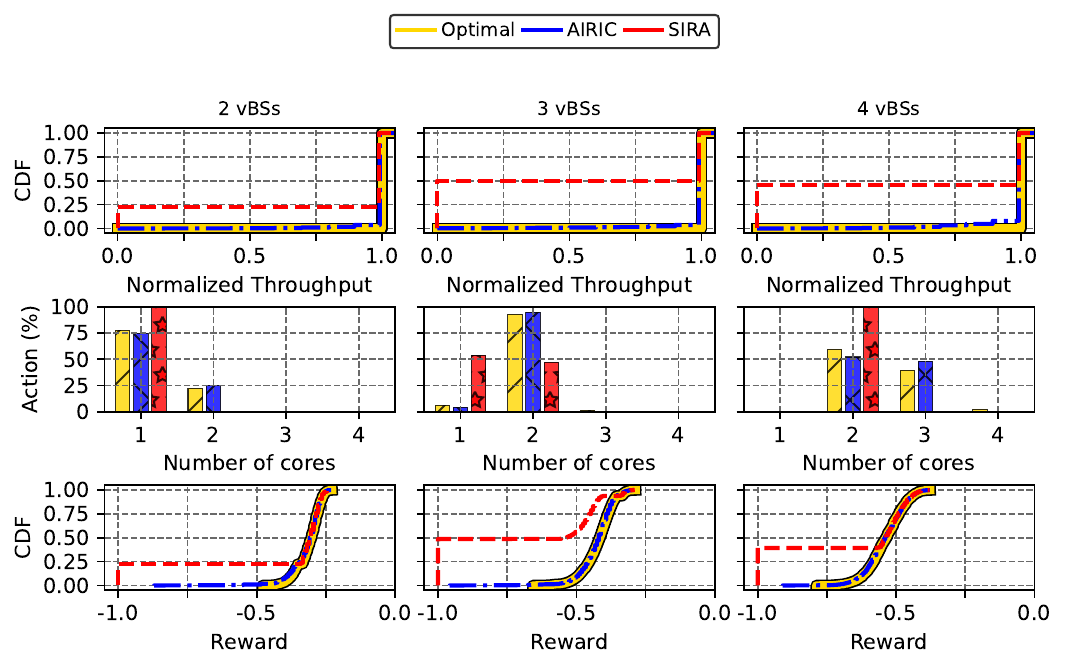}
 \vspace{-6mm}
 \caption{\small Performance benchmarking}
 \label{fig:eval:benchmark}
 \vspace{-4mm}
\end{figure}

\subsection{Performance benchmark}

To better understand the effectiveness of our solution, we now compare \name{} against a Single Instance Resource Allocation (SIRA) approach. SIRA is purposely designed to orchestrate optimal resources across vBS instances \emph{under the assumption of full computing isolation between instances}. Consequently, SIRA represents upper bounds attainable by existing works on vRAN CPU orchestration such as \cite{tripathi2023fair, vrain2}.
 
To evaluate \name{}, at every interval we choose uniformly at random the number of vBS instances, their DL/UL load and their DL/UL SNR, and use both approaches (\name{} and SIRA) to optimize the allocation of computing resources dynamically. In the case of SIRA, we use different (previously trained) models depending on the number of instances. For comparison, we also depict the performance of an oracle, labelled as ``Optimal'', that finds the optimal action offline by exhaustive search.

\begin{figure}[t!]
\centering
 \includegraphics[width=\columnwidth]{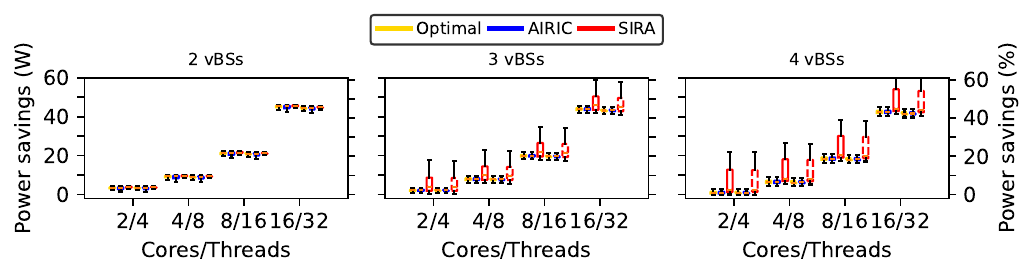}
 \vspace{-4mm}
 \caption{\small Power consumption savings}
 \label{fig:eval:power_savings}
 \vspace{-4mm}
\end{figure}

Fig.~\ref{fig:eval:benchmark} depicts the distribution of the normalized aggregate throughput performance of the system (top), the CPU assignments (middle), and the distribution of the reward achieved (bottom), for all the approaches conditioned to the presence of 2 (left), 3 (middle) and 4 (right) vBS instances. Conversely, Fig.~\ref{fig:eval:power_savings} depicts the absolute (left y-axis) and relative (right y-axis) power consumption savings achieved by all three approaches. These savings are in comparison to the power consumed when the default Linux scheduler manages all available CPU cores in the system, as indicated on the x-axis. The box plots represent the 25th and 75th percentiles (edges of the box), the median (line within the box), and the 5th-95th percentiles (error bars). We make three observations: The first observation is that \name{} provides substantial savings, comparable to the optimal benchmark. Perhaps surprisingly, SIRA shows mildly higher savings in some cases, which leads to our second observation: the savings provided by SIRA come at a huge price in throughput performance, as shown by Fig.~\ref{fig:eval:benchmark}. This is worse for denser scenarios: with 4 vBSs, SIRA barely saves $7\%$ computing resources more than \name{} in average but incurs $50\%$ throughput loss in exchange. This is due to the fact that SIRA ignores the additional computing overhead caused by the \rev{noisy neighbour problem} and often under-allocates resources, leading to PHY violations and throughput loss. The final observation is that \name{} provides a throughput performance that is remarkably close to that of ``Optimal''. Moreover,  Fig.~\ref{fig:eval:benchmark} (bottom) confirms that the reward distribution attained by \name{} is very close to the provided by the optimal oracle. These observations validate our design.

 \begin{figure}[t!]
  \centering
 \includegraphics[width=0.6\columnwidth]{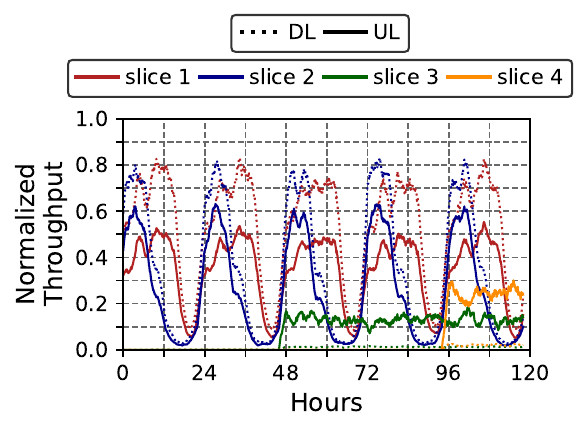}
 \vspace{-3mm}
 \caption{\small Realistic load traces}
  \label{fig:eval:traffic_traces}
  \vspace{-5mm}
 \end{figure}

\subsection{Realistic context traces}

We finally test \name{} with realistic context dynamics. To this end, we have generated context profiles for 4 different vBS instances, implementing \emph{network slices} with different context profiles, during 5 straight days. Fig.~\ref{fig:eval:traffic_traces} shows the time evolution of both DL and UL network load for these 4 traces. Slice 1 emulates the behavior of one eMBB vBS in the city center, with common diurnal load patterns. Slice 2 emulates a vBS serving an office building, with a peak load during office hours (9h - 17h). Both context dynamics are adapted from those in \cite{baysesian}. Slice 3 and 4, in turn, emulate IoT-serving vBSs with constant loads when they are operative. 

\begin{figure}[t!]
     \centering
     \begin{subfigure}[b]{0.49\columnwidth}
         \centering
         \includegraphics[width=\columnwidth]{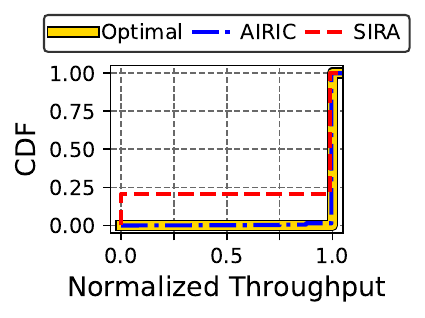}
         \vspace*{-3mm}
         \label{fig:eval:cdf_real}
     \end{subfigure}
     \hfill
     \begin{subfigure}[b]{0.49\columnwidth}
         \centering
         \includegraphics[width=\columnwidth]{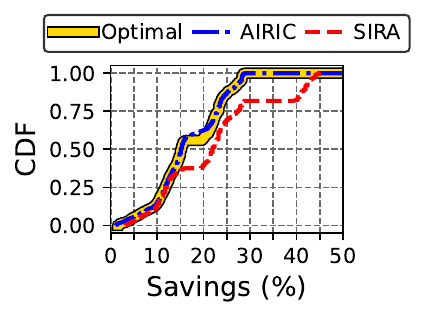}
         \vspace*{-3mm}
         \label{fig:eval:actionselec_real}
     \end{subfigure}
     \vspace{-6mm}
        \caption{Dynamic context profiles based on realistic traces.}
        \label{fig:eval:real}
    \vspace{-6mm}
\end{figure}

Fig.~\ref{fig:eval:real} depicts the distribution of the throughput performance (left) and the computing resource savings (right) of \name{}, SIRA and the optimal oracle. Like before, SIRA provides around $5\%$ higher CPU savings in average but incurs almost $25\%$ throughput loss over the 5 days as a consequence. Conversely, \name{} performs very closely to the oracle, with no throughput loss and around $17\%$ overall computing resource savings, which validates \name{} for realistic scenarios. 

\section{Related work}\label{sec:related}

\textbf{vRAN orchestration.}
There has been quite a number of pioneering work on the vRAN orchestration that embraces and builds upon the Open RAN paradigm to provide intelligent solutions on resource allocation for the deployment of vBSs over commercial off-the-shelf computing platform (e.g.,\cite{vrain, baysesian, edgebol, panchal2021enabling}) and provide energy-aware solutions (e.g., \cite{nuberu}) to optimize the energy consumption of underlying computing resources. For instance, \cite{panchal2021enabling} presents the implementation of a vBS capable of supporting URLLC slices. In the spectrum of computing resource allocation problems, the work of~\cite{baysesian} introduced a Bayesian learning model to optimize radio policies subject to hard power consumption constraints. EdgeBol~\cite{edgebol} proposed a non-real-time learning algorithm to optimize radio policies and non-radio service parameters jointly, and Concordia~\cite{concordia} addressed sharing computing resources with latency-elastic applications. 

RAN virtualization also enables sharing computing resources to reduce costs.
Making a decisive step forward towards cost-effective implementation of virtual and Open RAN, vrAIn~\cite{vrain} was the first work to jointly optimize the CPU allocation and radio policies for a given number vBSs deployment. More recently, \cite{tripathi2023fair} provided a solution to allocate computing resources among a vBS instance and a vertical service. They are considered as the most pioneer and relevant benchmark related to our work. But neither of the work look into and explore the \emph{noisy neighbor} problems caused by imperfect resource isolation over computing resources that are shared among virtual base stations, and no solution exist so far on computing required shared computing resources accounting for the impact of \rev{\emph{noisy neighbours} problem}, which is however significant on the vRAN performance, as pointed out in \S\ref{sec:intro} and as analyzed in \S\ref{sec:dissection}. Moreover, as shown in \S\ref{sec:evaluation}, this type of solutions requires independently-trained models depending on the total number of vBS deployed in the system. In contrast to all the prior work which does not support variable number of vBS instances, our approach learns the relationship between vBS instances and adapts naturally to different amount of instances over time. 

The \textit{noisy neighbor} problem in shared computing and networking environments has been extensively studied for cloud or container-based systems, but to the best of our knowledge, our work is the first to address this problem for vRAN shared computing platforms. In the following, we provide a sample of the most relevant contributions concerning isolation techniques, which are related to our analysis in \S\ref{sec:dissection}. 

\textbf{Network isolation.}
Noisy neighbor problems can be due to imperfect network traffic isolation. Different enforcement schemes have been proposed to ensure a high degree of traffic isolation among consolidated NFs, for instance, \cite{khalid2018iron} accounted for the time spent in the networking stack on behalf of a container, and \cite{tootoonchian2018resq} enhanced the cache isolation with careful sizing of I/O buffers, and \cite{panda2016netbricks} designed \textit{NetBricks} framework which embraced the zero-copy software isolation ideas. 

\textbf{Secure computing filters.} \emph{Seccomp}~\cite{seccomp} related work is mainly found in the computer security realm to harden security against attacks. \cite{manousis2020contention} proposed a reliable method to generate custom Seccomp profiles for arbitrary containerized applications to improve container security. ~\cite{9251949} proposed \textit{Draco} to address the lengthy rule-based checking programs against system calls and their arguments which lead to substantial execution overhead. And~\cite{10.1145/3474123.3486762} proposed \textit{Chestnut}, an automated approach for generating strict syscall filters of Seccomp with lower requirements and more restrictions.

\textbf{CPU isolation.}
Most work in this area is focused on advancing the CPU scheduling to prevent overheads caused by inter-core communication and context switching. For instance \cite{NetVM} developed a network packet processing platform built on top of the KVM platform and Intel DPDK library to support high-speed inter-VM communication through the scheduling VMs across different CPU cores. Besides, there are also some amount of work on exploring mapping of kernel thread partitioning techniques to CPU/GPU cores (e.g., \cite{10.1145/3126548}, \cite{10.5555/2616448.2616491}). 

\textbf{Cache memory isolation.}
One of the main causes of noisy neighbor problems is cache memory sharing, and more specifically the last-level cache (LLC). To address this problem, several works have proposed optimizing LLC partitioning and adopting Cache Allocation Technology (CAT) \cite{10.1145/2830555}\cite{10.1145/3062394}. In general, there are many different approaches to implement cache memory isolation, either by software (e.g., \cite{10.1145/3018113}) based on page coloring technique
or hardware (e.g., \cite{10.1145/2086696.2086723}) cache partitioning, or a combination of both (e.g., \cite{8091245}).
\section{Conclusions}

Contention for computing resources can jeopardize the performance and costs of virtualized radio access networks at scale as the number of base stations \emph{sharing} a computing platform grows. 
In our work, we have untangled the main sources for the increasing \emph{noisy neighbor} problem in vRANs (\emph{namespaces}, \emph{context switches}, \emph{security filters}, \emph{cache contention}) and quantified their relative impact towards the overall computing overhead. In order to address the identified noisy neighbor problem in vRANs, we have designed \name{}, which can adapt to varying contexts reconfiguring  computing platforms dynamically and achieving nearly the performance of an offline optimal oracle. Our results show that \name{} correctly dimensions the pool of computing cores and prevents vBSs from throughput collapse by accurately predicting the \rev{noisy neighbours problem}. \name{} leverages on a hybrid learning architecture comprising a Relation (RN) and a Deep \rev{Q-Network} (DQN) to  predict the best hardware configurations over time and counter the vRAN computing platforms \emph{sharing} negative effects; attaining over $99.9\%$ service availability and up to $30\%$ resource savings.  


\bibliographystyle{IEEEtran}
\bibliography{references}

\begin{IEEEbiography}[{\includegraphics[width=1in,height=1.25in,clip,keepaspectratio]{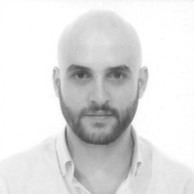}}]{Josep Xavier Salvat} received his Ph.D. from the Technical University of Kaiserlautern in 2022 and he currently works as senior research scientist in the 6G Network group at NEC Laboratories Europe, Heidelberg. He worked as reviewer of several international scientific conferences and journals, including IEEE Transactions on Mobile computing, IEEE ICC, and Computer Communications Journal and has  actively participated in several EU-founded projects, including H2020 5G-Crosshaul, H2020 5G-Transformer, and H2020 5Growth. His research interests lie in the application of machine learning to real-life computer communications systems, including resource allocation and energy efficiency problems.\end{IEEEbiography}

\begin{IEEEbiography}[{\includegraphics[width=1in,height=1.25in,clip,keepaspectratio]{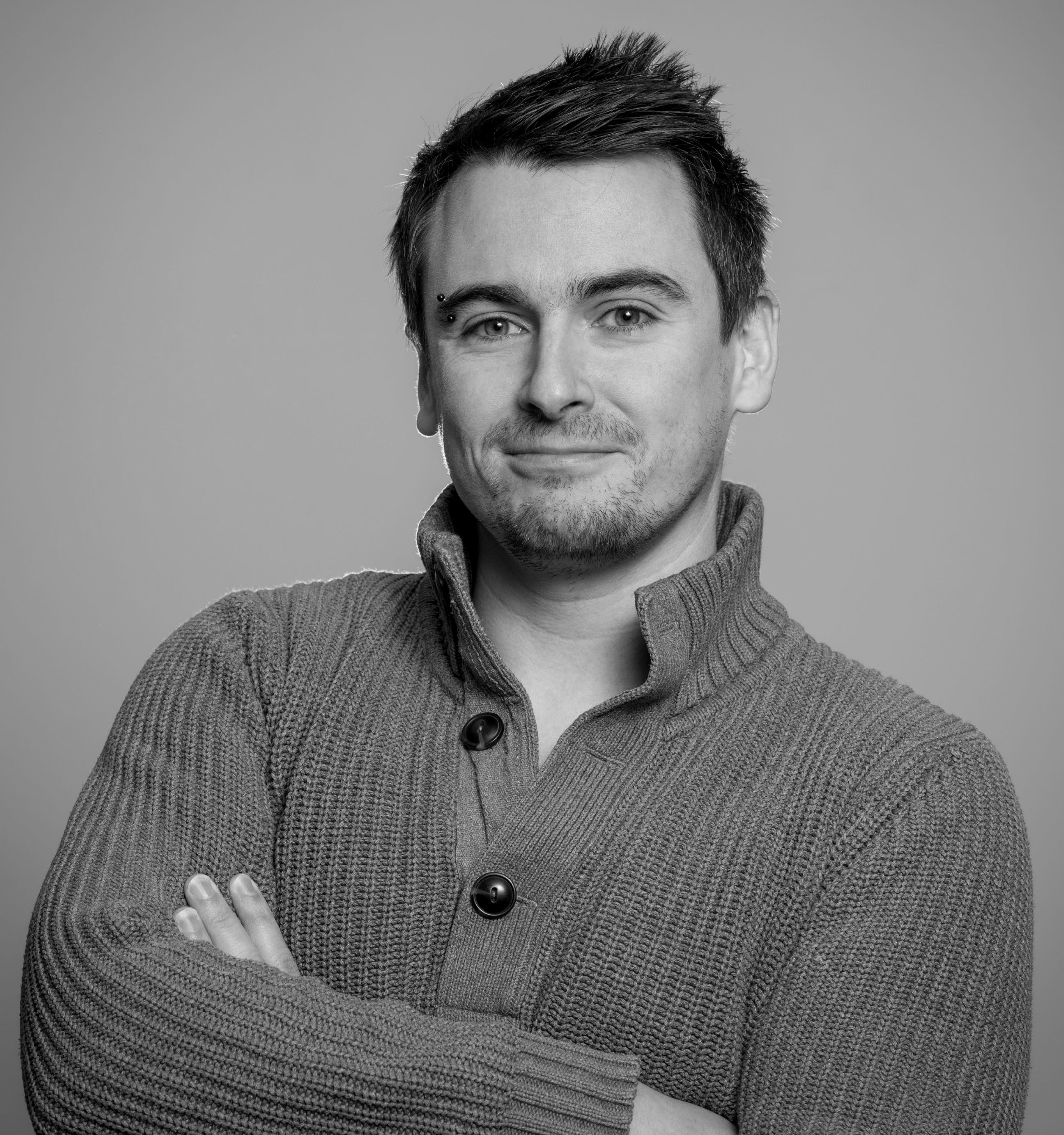}}]{Andres Garcia-Saavedra} is Principal Research Scientist at NEC Laboratories Europe.  Andres is a 5GPPP Technology Board member, served on the Program Committee and Editorial Team of several conferences and journals (such as IEEE ICC, or IEEE Transactions on Network Science and Engineering), published at top research venues (such as IEEE INFOCOM, or ACM MobiCom), and holds several patents. Andres received his PhD degree from the University Carlos III of Madrid (UC3M) in 2013. He then joined Trinity College Dublin (TCD), Ireland, as a research fellow. Since July 2015, he is with NEC Laboratories Europe. His research interests lie in the application of fundamental mathematics to real-life wireless communication systems. 
\end{IEEEbiography}

\begin{IEEEbiography}[{\includegraphics[width=1in,height=1.25in,clip,keepaspectratio]{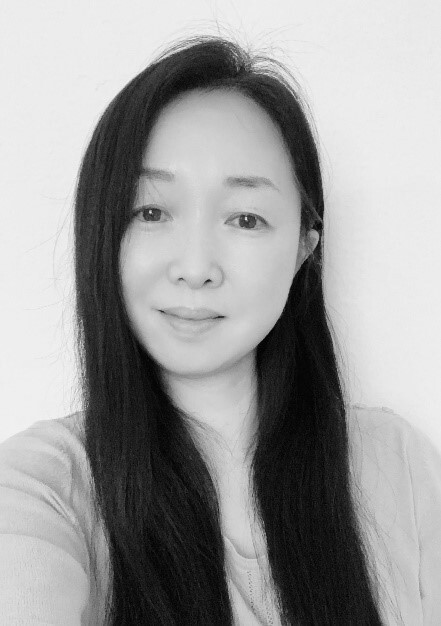}}]{Xi Li} is a Senior Researcher on 6G Networks R\&D at NEC Laboratories Europe, Germany, and the Vice Chairman of the 5GPPP Architecture Working Group. She received her M.Sc. in 2002 from the Technical University of Dresden and Ph.D. in 2009 from University of Bremen, Germany. She is currently the Technical Manager of the EU H2020 5Growth project and from 2015-2019 she has led technical work package in EU H2020 5G-Crosshaul and 5G-TRANSFORMER projects.  Previously, she was a senior researcher fellow and lecturer at the University of Bremen and a solution designer at Telefonica, Germany. She has published 80+ journal and conference publications, and given many invited talks in various industrial events and international conferences. She is an inventor of 18 patents including 7 granted patents, and active in contributing to IETF CCAMP WG with two published RFCs and received best overall award at IETF’99 Hackathon in 2017. Her research interests comprise the design for next generation mobile and wireless networks, open and virtualized RAN, distributed edge platform solutions, applying AI/ML for resource and service management and automation. 
\end{IEEEbiography}

\begin{IEEEbiography}[{\includegraphics[width=1in,height=1.25in,clip,keepaspectratio]{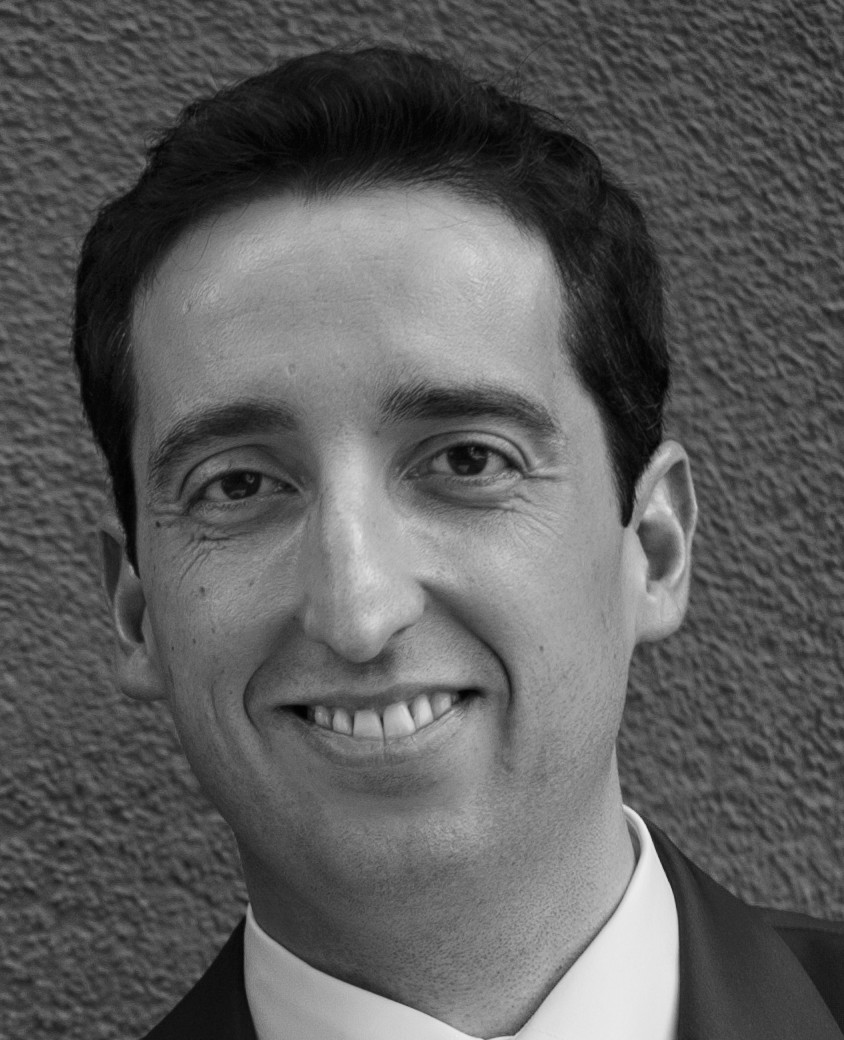}}]{Xavier Costa-P\'erez} (M'06--SM'18) is Scientific Director at the i2Cat R\&D Center, Head of Beyond 5G Networks R\&D at NEC Laboratories Europe and Research Professor at ICREA. His team contributes to products roadmap evolution as well as to European Commission R\&D collaborative projects and received several awards for successful technology transfers. In addition, the team contributes to related standardization bodies: 3GPP, O-RAN, ETSI RIS and IETF. Xavier has been a 5GPPP Technology Board member, served on the Program Committee of several conferences (including IEEE Greencom, WCNC, and INFOCOM), published at top research venues and holds several patents. He also serves as Editor of IEEE Transactions on Mobile Computing and Transactions on Communications journals. He received both his M.Sc. and Ph.D. degrees in Telecommunications from the Polytechnic University of Catalonia (UPC) in Barcelona and was the recipient of a national award for his Ph.D. thesis.
\end{IEEEbiography}

\end{document}